\documentclass[12pt, reqno]{amsart}
\usepackage{mathrsfs}
\usepackage{amsmath}
\usepackage{amsxtra}
\usepackage{amscd}
\usepackage{amsthm}
\usepackage{amsfonts}
\usepackage{amssymb}
\usepackage{pstricks}
\usepackage{pstricks,pst-node}

\theoremstyle{definition}

\theoremstyle{remark}

\numberwithin{equation}{section}

\newcommand{\Eq}{{\mathcal Eq}}

\begin{document}


\title[Photon Rockets]{Bondi-Sachs metrics and Photon Rockets}

\author{Huabin Ge}
\address[Huabin Ge]{School of Mathematical Sciences,Peking University,Beijing 100871,PR China}
\email{gehuabin@pku.edu.cn}

\author{Mingxing Luo}
\address[Mingxing Luo]{Zhejiang Institute of Modern Physics, Department of Physics, Zhejiang
University, Hangzhou, Zhejiang 310027, PR China}
\email{luo@zimp.zju.edu.cn}

\author{Qiping Su}
\address[Qiping Su]{Department of Physics, Hangzhou Normal University, Hangzhou, 310036, China}
\email{sqp@hznu.edu.cn}

\author{Ding Wang}
\address[Ding Wang, Xiao Zhang]{Institute of Mathematics,
Academy of Mathematics and Systems Science, Chinese Academy of
Sciences, Beijing 100190, China}
\email{wangding@amss.ac.cn, xzhang@amss.ac.cn}

\author{Xiao Zhang}

\begin{abstract}
We study the Bondi-Sachs rockets with nonzero cosmological constant. We observe that the acceleration
of the systems arises naturally in the asymptotic symmetries of (anti-) de Sitter spacetimes.
Assuming the validity of the concepts of energy and mass previously introduced in asymptotically flat spacetimes,
we find that the emission of pure radiation energy balances the loss of the Bondi mass in certain special families
of the Bondi-Sachs rockets, so in these there is no gravitational radiation.
\\\\
PACS(2010): 04.30-w, 04.20Ha
\end{abstract}


\maketitle

%
%
\section{Introduction}\label{introd}
In \cite{Kin69}, Kinnersley introduced and studied a pure radiation Robinson-Trautman spacetime of algebraic type D. This spacetime
provides a model for a particle accelerating with the emission of photon fluid and is interpreted as a ``photon
rocket'' in physics (e.g. \cite{SKMHH03}). Surprisingly, Bonnor \cite{Bonnor94} found that the Kinnersley rocket
does not lose energy with respect to the flat background frame. This absence of gravitational radiation was
later analyzed by Damour \cite{Damour94} using the post-Minkowskian perturbation method. In this case the linearized
gravitational waves are generalized by the motion of the massive point-like rocket and by the energy-momentum
distribution of the photon fluid, where the two contributions cancel each other. Bonnor \cite{Bonnor96} also obtained
a family of Kinnersley-type rockets of algebraic type II, which belongs to axi-symmetric pure radiation Robinson-Trautman spacetimes. In \cite{vonderGonna98}, von der G\"{o}nna and Kramer showed that the Kinnersley family of type D is the only axi-symmetric and asymptotically flat Robinson-Trautman spacetime with pure radiation but no gravitational radiation. In \cite{Cornish00}, Cornish proved that pure radiation Robinson-Trautman spacetimes with zero news function must be the Kinnersley photon rocket metrics.

In all these works the cosmological constant was zero. However, recent cosmological observations indicated
that our universe has a positive cosmological constant. It becomes an important question to study the relevant problems in the case of positive cosmological constant (e.g. \cite{Luo} and references therein). In \cite{Podolsky08}, Podolsk\'{y} studied the family of Kinnersley's rockets and Bonnor's rockets with nonzero cosmological constant. These spacetimes are axi-symmetric Robinson-Trautman type which describe photon rockets moving arbitrarily
in de Sitter or anti-de Sitter universe due to anisotropic emission of pure radiation. He also analyzed the trajectories of the photon rockets which accelerate in the corresponding Minkowski or (anti-)de Sitter spacetimes.

As the Bondi-Sachs spacetimes represent one of the most important radiation spacetimes, it is our interest to
investigate the photon rocket spacetimes of Bondi-Sachs type with nonzero cosmological constant. By applying asymptotic analysis to the Bondi-Sachs spacetimes with nonzero cosmological constant, we observe that certain asymptotic functions $B$, $X$ and $Y$ can't be reduced to zero by coordinate transformations in general. This is completely different from the case of zero cosmological constant. Interestingly, the acceleration of the systems arises naturally in the appearance of certain asymptotic function (Section 2). We then obtain a family of exact pure radiation Bondi-Sachs spacetimes with cosmological constant, which we refer as the Bondi-Sachs rockets (Section 3). And we study the emission of pure radiation energy and the loss of the Bondi mass for these rockets. We find, in two families of rockets either $B=0$
or axi-symmetric spacetimes with $B=\frac{1}{2}\ln (C+\cos\theta)$ for some constant $C>1$, the emission of pure radiation energy balances the loss of the Bondi mass. And no gravitational radiation occurs in these cases (Section 4).

\section{Bondi-Sachs Metrics}\label{section-bondi-sachs}

Suppose a spacetime has a family of non-intersecting null hypersurfaces given by the level
sets of smooth function $u$. We choose coordinates $x^0 =u$, $x^1=r$, $x^2=\theta$ and $x^3=\phi$
where $r$ is a luminosity distance along the null rays. Fixing $u$, $\theta$
and $\phi$ and letting $r$ vary, we obtain a null ray with the tangent vector
$\frac{dx^a}{dr}=\delta ^a _1$. As $l _a =\partial _a u =\delta ^0 _a$
is the normal covariant vector field to these null hypersurfaces which is also tangent
to the null hypersurfaces, we must have $\lambda \delta ^a_1 =l^a =g^{ab}\partial _b u =g^{a0}$.
Therefore
$
g^{00}=g^{02}=g^{03}=0.
$
These conditions are equivalent to the conditions
$
g_{11}=g_{12}=g_{13}=0.
$
Under these conditions, the spacetime takes the Bondi-Sachs metric
\cite{Bondi,Sachs,Burg} even if the cosmological constant $\Lambda$ is
nonzero:
\begin{equation}\label{ds2}
ds^2=\left(e^{2\beta}
\frac{V}{r}-r^2h_{AB}U^AU^B\right)du^2+2e^{2\beta}dudr+2r^2h_{AB}U^Bdudx^A-r^2h_{AB}dx^Adx^B
\end{equation}
where $A,B=2,3$, $U^2= U$, $U^3= W\csc\theta$,
\[
h\equiv\left(
\begin{array}{ll}
 h_{22}  & h_{23}  \\
 h_{32}  & h_{33}
\end{array}
\right) \equiv\left(
\begin{array}{ll}
 e^{2\gamma}\cosh 2\delta  & \sinh 2\delta\sin\theta  \\
 \sinh 2\delta\sin\theta  & e^{-2\gamma}\cosh 2\delta\sin^2\theta
\end{array}
\right),
\]
$\beta$, $\gamma$, $\delta$, $V$, $U$, $W$ are functions of $u$, $r$
and points on unit 2-sphere parameterized by  $\theta$, $\phi$, i.e.,
they take the same values at $\phi =0$ and $\phi=2\pi$. Let
$\Omega _{ij}=R_{ij}+\Lambda g_{ij}$, $\Omega=g^{ij}\Omega _{ij}$.

Now we study the vacuum Einstein field equations
\begin{eqnarray}
\Omega _{ij}=0  \label{vacuum}
\end{eqnarray}
for the metric (\ref{ds2}) with cosmological constant
$\Lambda$. These ten equations break up into three groups,
as in the case of zero cosmological constant \cite{Bondi,Sachs,Burg}:
\begin{itemize}
\item[(A)] six main equations
\[
\Omega_{11}=\Omega_{12}=\Omega_{13}=\Omega_{22}=\Omega_{23}=\Omega_{33}=0;
\]
\item[(B)] one trivial equation
\[
\Omega_{01}=0;
\]
\item[(C)] three supplementary equations
\[
\Omega_{00}=\Omega_{02}=\Omega_{03}=0.
\]
\end{itemize}

The Bianchi identity implies that
\begin{eqnarray}
\nabla ^i \left(\Omega_{ij}-\frac{\Omega}{2}g_{ij}\right)=0 \Longleftrightarrow
g^{ij}\left(\Omega_{ki ,j} -\frac{1}{2}\Omega_{ij, k} - \Gamma^l_{ij}\Omega_{kl}\right)=0. \label{Bianchi}
\end{eqnarray}

If the main equations hold everywhere, then (\ref{Bianchi}) implies
that
\begin{itemize}
\item[(i)] the trivial equation holds everywhere,
\item[(ii)] the supplementary equations hold everywhere if they hold on a
hypersurface $r=r_0$ for some constant $r_0 >0$.
\end{itemize}
Indeed, using the explicit formulas of the Christoffel symbols
\cite{Sachs}, we find
$
g^{ij }\Gamma_{ij}^0=-2e^{-2\beta}/r.
$
Supposing the six main equations hold, we have
\[
\begin{aligned}
k=1: & \;(\ref{Bianchi}) \Longrightarrow g^{ij}\Gamma^0_{ij}\Omega_{01}=0 \Longrightarrow
\Omega_{01}=0;\\
k=2: & \;(\ref{Bianchi}) \Longrightarrow g^{01}\Omega_{02,r}-g^{ij}\Gamma^0_{ij}\Omega_{02}=0 \Longrightarrow
r^{-2}e^{-2\beta} \left(r^2\Omega_{02}\right)_{,r}=0;\\
k=3: & \;(\ref{Bianchi}) \Longrightarrow g^{01}\Omega_{03,r}-g^{ij}\Gamma^0_{ij}\Omega_{03}=0 \Longrightarrow
r^{-2}e^{-2\beta} \left(r^2\Omega_{03}\right)_{,r}=0;
\end{aligned}
\]
If the three supplementary equations hold on a hypersurface $r=r_0$, we get
\[
\Omega_{02}=\Omega_{03}=0.
\]
Using the Bianchi identities with $k=0$ together with the
above two equations, we obtain
\[
r^{-2}e^{-2\beta} \left(r^2 \Omega_{00}\right)_{,r}=0 \Longrightarrow
\Omega_{00}=0.
\]

The six main equations can break up further into two groups:
\begin{itemize}
\item[(A1)] four hypersurface equations
\[
\Omega_{11}=\Omega_{12}=\Omega _{13}=0\Longleftrightarrow R_{11}=R_{12}=R _{13}=0,
\]
\[
\begin{aligned}
&\left(e^{-2\gamma} \Omega _{22}+e^{2\gamma} \csc^2\theta
\Omega _{33}\right)\cosh2\delta-2\csc\theta \Omega _{23}\sinh2\delta=0\\
\Longleftrightarrow &\left(e^{-2\gamma} R _{22}+e^{2\gamma} \csc^2\theta
R _{33}\right)\cosh2\delta-2\csc\theta R _{23}\sinh2\delta=2 \Lambda r^2;
\end{aligned}
\]
\item[(A2)] two standard equations
\[
e^{-2 \gamma} \Omega _{22}-e^{2 \gamma}\csc ^2\theta  \Omega _{33}=0
\Longleftrightarrow e^{-2 \gamma} R _{22}-e^{2 \gamma}\csc ^2\theta  R _{33}=0,
\]
\[
\begin{aligned}
&\left(e^{-2\gamma} \Omega _{22}+e^{2\gamma} \csc^2\theta
\Omega _{33}\right)\sinh2\delta-2\csc\theta \Omega _{23}\cosh2\delta=0\\
\Longleftrightarrow &
\left(e^{-2\gamma} R _{22}+e^{2\gamma} \csc^2\theta
R _{33}\right)\sinh2\delta-2\csc\theta R _{23}\cosh2\delta=0.
\end{aligned}
\]
\end{itemize}

In terms of the explicit expressions of the left hand sides of the above equations \cite{Burg},
we can write the main equations as $\Eq (i)=0$, $1 \leq i \leq 6$, where $\Eq (i)$ are
given in the appendix.

In \cite{Smalley, Chrusciel},  Smalley, Chru\'sciel analyzed asymptotic Bondi's axi-symmetric
spacetimes with positive cosmological constant. Now we will apply asymptotic analysis to
more general Bondi-Sachs' spacetimes. We study the vacuum field equations (\ref{vacuum}) for
the Bondi-Sachs metric (\ref{ds2}) which is asymptotically de Sitter. For this purpose, we
assume that $\gamma$ and $\delta$ have the following asymptotic expansions
\[
\begin{aligned}
\gamma&=\frac{c}{r}+\Big(-\frac{1}{6} c^3-\frac{3}{2} d^2 c+C\Big)
   \Big(\frac{1}{r}\Big)^3+O\Big(\frac{1}{r}\Big)^4,\\
\delta&=\frac{d}{r}+\Big(-\frac{1}{6}d^3+\frac{1}{2} c^2d+D\Big)
   \Big(\frac{1}{r}\Big)^3+O\Big(\frac{1}{r}\Big)^4.
\end{aligned}
\]
However, as pointed out by Chru\'sciel \cite{Chrusciel}, the ``asymptotically de Sitter'' condition allows
much more general forms in the expansions of $\gamma$ and $\delta$, and he interpreted, in the axi-symmetric case,
the above expansion of $\gamma$ with the additional component independent of $r$ as a ``radiation condition". In this
paper, we consider only the simpler case that the components independent of $r$ vanish. Now,
using the equation $\Eq (1)$, we obtain
\[
\begin{aligned}
\beta =B -\frac{1}{4} \Big(c^2+d^2\Big)
\Big(\frac{1}{r}\Big)^2+\frac{1}{8} \Big(c^4+2 c^2 d^2 +d^4-6 c C-6 d D\Big)
   \Big(\frac{1}{r}\Big)^4+O\Big(\frac{1}{r}\Big)^5.
\end{aligned}
\]
Using $\Eq (2)=\Eq (3)=0$, we obtain
\[
\begin{aligned}
W=&X+2 e^{2 B} B_{,\phi} \csc \theta\frac{1}{r}+e^{2 B} \Big(2 c
   B_{,\phi}\csc \theta+c_{,\phi}\csc \theta- 2 d\cot \theta +2 d B_{,\theta}-d_{,\theta}\Big) \Big(\frac{1}{r}\Big)^2\\
   &+e^{2 B} \Big( B_{,\phi} c^2 \csc \theta+3c c_{,\phi}\csc \theta
   -2c d_{,\theta} +2 P+ d^2 B_{,\phi} \csc \theta +3 d d_{,\phi} \csc \theta\\
   &+2 c_{,\theta} d \Big) \Big(\frac{1}{r}\Big)^3+O\Big(\frac{1}{r}\Big)^4,\\
U=&Y+2 e^{2 B} B_{,\theta}\frac{1}{r}-e^{2 B} \Big( 2 d B_{,\phi}\csc \theta+d_{,\phi}\csc \theta+ 2 c\cot \theta +2 cB_{,\theta}+c_{,\theta}\Big)\Big(\frac{1}{r}\Big)^2\\
   &+e^{2 B}\Big( c^2 B_{,\theta}+4 c^2 \cot \theta+2 c d_{,\phi} \csc \theta+3 c c_{,\theta} +2 N -2 d c_{,\phi} \csc \theta +4 d^2 \cot \theta \\ &+d^2 B_{,\theta}+3 d d_{,\theta}\Big)\Big(\frac{1}{r}\Big)^3 +O\Big(\frac{1}{r}\Big)^4.
\end{aligned}
\]
Substitute these into $\Eq (4)=0,$ we obtain
\begin{eqnarray*}
\begin{aligned}
V=&-\frac{e^{2 B} \Lambda }{3}r^3+\Big(\cot
\theta Y+\csc \theta
X_{,\phi}+Y_{,\theta}\Big)r^2+e^{2 B} \Big( 4 B_{,\phi}^2 \csc ^2\theta+2 B_{,\phi \phi} \csc ^2\theta\\
&+\frac{1}{2}\Lambda c^2+\frac{1}{2}\Lambda d^2+ 2 B_{,\theta} \cot \theta+4 B_{,\theta}^2+2 B_{,\theta \theta}+1\Big)r-2 M+O\Big(\frac{1}{r}\Big).
\end{aligned}
\end{eqnarray*}
In the above formulas, $B(u,\theta, \phi)$, $c(u,\theta, \phi)$, $C(u,\theta, \phi)$, $D(u,\theta, \phi)$, $d(u,\theta, \phi)$,
$X(u,\theta, \phi)$, $Y(u,\theta, \phi)$, $N(u,\theta, \phi)$, $P(u,\theta, \phi)$ and $M(u,\theta, \phi)$ are functions of $u$, $\theta$, $\phi$ which are defined on unit 2-sphere parameterized by  $\theta$, $\phi$ for fixed $u$. Now, we have
\begin{eqnarray*}
\begin{aligned}
\Eq (5)=&-\frac{1}{6} \Big(-2 e^{2 B} \Lambda c-3 \cot \theta Y-3 \csc\theta X_{,\phi}+3Y_{,\theta}\Big)\\
        &-\frac{1}{3} d^2 \Big(2 e^{2 B} \Lambda  c+3 \cot \theta Y+3 \csc \theta X_{,\phi}-3 Y_{,\theta}\Big)
          \Big(\frac{1}{r}\Big)^2+O\Big(\frac{1}{r}\Big)^3=0,\\
\Eq (6)=&-\frac{1}{6} \Big(-2 e^{2 B} \Lambda d-3 \cot \theta X+3 \csc\theta Y_{,\phi}+3X_{,\theta}\Big)\\
        &+\frac{c}{3} \Big( 2 e^{2 B} \Lambda c d +3c\left(-\cot \theta X+\csc \theta Y_{,\phi}+X_{,\theta}\right)\\
        &+6 d \left(\cot\theta Y+\csc \theta X_{,\phi}-Y_{,\theta}\right)\Big) \Big(\frac{1}{r}\Big)^2+O\Big(\frac{1}{r}\Big)^3=0.
\end{aligned}
\end{eqnarray*}
Let $X=\sin \theta \bar X$, $Y=\sin \theta \bar Y$, and $\triangle =\partial ^2 _\theta +\cot\theta \partial _\theta + \csc^2 \theta \partial ^2 _\phi$ the Laplacian operator on $S^2$. Then the above equations give that
\begin{eqnarray*}
\bar X_{,\phi} -\sin\theta \bar Y_{,\theta} =-\frac{2}{3}\Lambda e^{2 B} c,\quad
\sin \theta \bar X_{,\theta}+ \bar Y_{,\phi} =\frac{2}{3}\Lambda e^{2 B} d.
\end{eqnarray*}
Thus $\bar X$, $\bar Y$ satisfy the following equations
\begin{eqnarray}
&&\triangle \bar X = \frac{2}{3}\Lambda \Big(e^{2 B} d\Big)_{,\theta} \csc \theta-\frac{2}{3}\Lambda \Big(e^{2 B} c \Big)_{,\phi} \csc^2 \theta \label{eq-X}\\
&&\triangle \bar Y = \frac{2}{3}\Lambda \Big(e^{2 B} c\Big)_{,\theta} \csc \theta+\frac{2}{3}\Lambda \Big(e^{2 B} d \Big)_{,\phi} \csc^2 \theta.\label{eq-Y}
\end{eqnarray}
Now we conclude that, if the following conditions hold
\begin{eqnarray}
\big(e^{2B} c\big)(u, 0, \phi)=\big(e^{2B} c\big)(u, \pi, \phi),\quad \big(e^{2B} d\big)(u, 0, \phi)=\big(e^{2B} d\big)(u, \pi, \phi),
\end{eqnarray}
then there exist $\bar X$, $\bar Y$, equivalently, $X$, $Y$, satisfying (\ref{eq-X}), (\ref{eq-Y})
which are unique up to certain functions $\rho(u)$, $\sigma(u)$ depending on $u$ only. Therefore we can use the power series to solve the
field equations. Interestingly, $\sigma(u)$ results in the acceleration of the systems. Moreover, $\sigma(u)$, $\rho(u)$ result in the variation of the shear.

If the cosmological constant $\Lambda =0$, then $X$, $Y$ must be zero, and $B$ can be chosen as zero
by suitable coordinate transformation \cite{Bondi,Sachs,Burg}. However, when $\Lambda \neq 0$, this
is not true in general. For instance, we study the case of Bondi's axi-symmetric spacetime,
i.e., the functions are independent on $\phi$.
Choosing the following coordinate transformation
\begin{eqnarray*}\label{coortr}
\begin{cases}
\begin{aligned}
u&=u_0(\bar u, \bar \theta) +\frac{u_1(\bar u, \bar \theta) }{\bar r }+O\Big(\frac{1}{\bar r }\Big)^2,\\
r&=\bar r+r_0(\bar u, \bar \theta)+\frac{r_1(\bar u, \bar \theta) }{\bar r }+O\Big(\frac{1}{\bar r }\Big)^2,\\
\theta&=\bar \theta +\frac{\theta_1(\bar u, \bar \theta) }{\bar r }+O\Big(\frac{1}{\bar r }\Big)^2,\\
\phi&=\bar \phi,
\end{aligned}
\end{cases}
\end{eqnarray*}
we obtain
\[
\begin{aligned}
\bar g_{11}=&-\Big(\frac{1}{3} e^{4 B } \Lambda u_1 ^2 +2 e^{2 B }u_1 +(\theta_1 -u_1  Y )^2\Big)
   \Big(\frac{1}{\bar r }\Big)^2+O\Big(\frac{1}{\bar r }\Big)^3,\\
\bar g_{12}=&e^{2 B } u_{0, \bar \theta}+\frac{1}{3} e^{4 B } \Lambda u_1
   u_{0,\bar \theta}+(u_1  Y -\theta_1 ) (Y u_{0,\bar \theta}-1)+O\Big(\frac{1}{\bar r }\Big),\\
\bar g_{33}=&-\bar r^2 \sin ^2\bar \theta +O\big(\bar r\big),\\
\bar g_{22} \bar g_{33}=&\bar r^4\sin ^2 \bar \theta  \Big(\frac{1}{3} e^{4 B } \Lambda u_{0,\bar \theta}^2+(u_{0,\bar \theta} Y-1)^2\Big)\\
&+\bar r^3 \sin  \bar \theta \Big(\frac{2}{3} e^{4 B } \Lambda u_{0,\bar \theta} \big((\cos  \bar \theta  \theta_1 -\sin  \bar \theta  c
   +2 \sin  \bar \theta r_0) u_{0,\bar \theta}+\frac{1}{3}\sin  \bar \theta  u_{1,\bar \theta}\big)\\
   &+2(u_{0,\bar \theta}Y-1) \big((\cos  \bar \theta  u_1 +2 \sin  \bar \theta r_0) (u_{0,\bar \theta}Y-1)
   +\sin  \bar \theta (u_{1,\bar \theta}Y-\theta_{1,\bar \theta})\big)\\
   &- e^{2 B } \sin  \bar \theta u_{0,\bar \theta} \big(4 B_{,\theta} +u_{0,\bar \theta} (Y  \cot \theta -4 Y B_{,\theta} +Y_{,\theta} )\big)\Big)+O(\bar r^2),\\
\bar g_{01}=&\Big(e^{2 B }-\theta_1  Y +u_1 \big(Y ^2+\frac{1}{3} e^{4B} \Lambda \big)\Big) u_{0,\bar u} +O\Big(\frac{1}{\bar r }\Big).
\end{aligned}
\]
Using $\bar g_{22} \bar g_{33}=\bar r^4 \sin ^2 \bar \theta$, we obtain
\begin{eqnarray}
u_{0,\bar \theta}=0 \ \text{or} \ \frac{6 Y }{3 Y ^2+e^{4 B } \Lambda }.\label{u02}
\end{eqnarray}
Using $\bar g_{11}=\bar g_{12}=0$, we obtain
$
\bar g_{01}=e^{2 B } u_{0,\bar u} +O\Big(\frac{1}{\bar r }\Big).
$
Thus, in order to make $B$ zero, we must choose
\begin{eqnarray}
u_{0,\bar u} =e^{-2B}. \label{u00}
\end{eqnarray}
So $u_0$ can be solved and {\it B can be reduced to zero if $B=B(u)$}. In general, one could not solve (\ref{u02}), (\ref{u00}) to find coordinates such that
$B$ and $Y$ are simultaneously zero.

\section{Bondi-Sachs Rockets}\label{sect-BS-roc}

In this section, we shall discuss the pure radiation photon rocket solutions of the Bondi-Sachs metric (\ref{ds2})
\begin{eqnarray}
\Omega _{ij}=\frac{\eta^2(u, \theta, \phi)}{r^2} \partial_i u \partial_j u  \label{PhotonRoc}
\end{eqnarray}
under the conditions $\gamma=\delta=0$. These conditions imply that the spacelike 2-spheres lying in the hypersurfaces
$u=constant$ and $r=constant$ take the standard metric $-r^2 (d\theta^2+\sin^2 \theta d \phi ^2)$. Using the equations $\Eq (i)=0$, $i=1, \cdots, 4$, we obtain
\[
\begin{aligned}
\beta =&B,\quad W=\frac{2 e^{2 B} P}{r^3}+X+\frac{2 e^{2 B} \csc \theta B_{,\phi}}{r},\quad U=\frac{2 e^{2 B} N}{r^3}+Y+\frac{2 e^{2 B} B_{,\theta}}{r},\\
V=&-\frac{1}{3} e^{2 B} \Lambda  r^3+\Big(\cot \theta Y+\csc \theta X_{,\phi}+Y_{,\theta}\Big) r^2\\
&+e^{2 B} \Big( 4 B_{,\phi}^2 \csc ^2\theta+2 B_{,\phi \phi} \csc ^2\theta+2 B_{,\theta} \cot \theta+4 B_{,\theta}^2+2 B_{,\theta \theta}+1\Big) r\\
&-2 M -e^{2 B}\Big(P_{,\phi} \csc \theta -4 P B_{,\phi} \csc \theta + N \cot \theta-4 N B_{,\theta}+N_{,\theta}\Big)\frac{1}{r}\\
&+3 e^{2 B} \Big(N^2+P^2\Big)\Big(\frac{1}{r}\Big)^3.
\end{aligned}
\]
Now, we have
\[
\begin{aligned}
\Eq (5)=&\frac{9 e^{2 B} \left(N^2-P^2\right)}{2 r^5}-\frac{e^{2 B} \left( 8
   P B_{,\phi}\csc \theta +P_{,\phi}\csc \theta + N\cot \theta-8N B_{,\theta}-N_{,\theta}\right)}{2 r^3}\\
   &-\frac{1}{2} \left(-\cot \theta Y-\csc \theta X_{,\phi}+Y_{,\theta}\right)=0,\\
\Eq (6)=&\frac{9 e^{2 B} N P}{r^5}+\frac{e^{2 B} \left(8 N B_{,\phi}\csc \theta +N_{,\phi}\csc \theta - P\cot \theta-8P B_{,\theta}+P_{,\theta}\right)}{2 r^3}\\
&-\frac{1}{2} \left(-\cot \theta X+\csc \theta Y_{,\phi}+X_{,\theta}\right)=0.
\end{aligned}
\]
Therefore, $N$=$P$=0. As $c$=$d$=0, the equations (\ref{eq-X}), (\ref{eq-Y}) imply that $\bar X$, $\bar Y$ are harmonic functions
over $S^2$, which are therefore functions depending on $u$ only. Thus $X=\sin \theta  \rho(u)$, $Y=\sin \theta  \sigma(u)$.
Now the supplementary equations $r^2 \Omega_{02}=r^2 \Omega_{03}=0$ give $4 M B_{,\theta}+M_{,\theta}=0$, $4 M B_{,\phi}+M_{,\phi}=0$.
Solving these equations, we obtain
\begin{equation}\label{lastequ0}
M=\tau(u) e^{-4B}
\end{equation}
for some arbitrary function $\tau(u)$.  Furthermore, that $r^2 \Omega_{00}=\eta^2$ gives
\begin{equation}\label{lastequ}
\begin{aligned}
&e^{4 B}
   \Big(4 B_{,\phi \phi}^2 \csc ^4\theta +B_{,\phi \phi \phi \phi} \csc ^4\theta
   -2 \cot \theta  B_{,\theta \phi \phi} \csc ^2\theta -B_{,\theta \theta} \csc ^2\theta \\
&+8 B_{,\phi} ^2 \left(2 B_{,\phi \phi} \csc ^2\theta
   +2 \cot \theta
   B_{,\theta} +2 B_{,\theta \theta}+1\right) \csc ^2\theta \\
&+4 B_{,\phi \phi} \left(\csc ^2\theta
   +4 B_{,\theta} ^2-2 \cot \theta  B_{,\theta} +2 B_{,\theta \theta}\right) \csc ^2\theta\\
&+8 B_{,\phi}
   \left(B_{,\phi \phi \phi} \csc ^2\theta +\cot \theta  B_{,\theta \phi}
   +B_{,\theta \theta \phi}\right) \csc ^2\theta +2
   B_{,\theta \theta \phi \phi} \csc ^2\theta\\
&+16 \cot \theta  B_{,\theta} ^3+4 B_{,\theta \theta}^2-4 B_{,\theta} ^2
   \left(\cot ^2\theta -4 B_{,\theta \theta}\right)+B_{,\theta \theta}
   +2 \cot \theta  B_{,\theta \theta \theta}\\
&+B_{,\theta}  \big(8 B_{,\theta \phi \phi} \csc ^2\theta+\cot \theta  \left(\csc ^2\theta
   +16 B_{,\theta \theta}+2\right)+8 B_{,\theta \theta \theta}\big)+B_{,\theta \theta \theta \theta}\Big)\\
&+4 M B_{,u}-2 M_{,u}-6 \cos \theta  M \sigma-3\sin \theta M_{,\theta} \sigma -3 \rho M_{,\phi}=\eta ^2.
\end{aligned}
\end{equation}

If $B$ solves the equation (\ref{lastequ}) where $M$ is given by the equation (\ref{lastequ0}),
then the following Bondi-Sachs metrics will solve the field equations (\ref{PhotonRoc})
\begin{equation}\label{BS-roc}
\begin{aligned}
ds^2=&e^{2B}\frac{V}{r} du^2 +2e^{2B}dudr -r^2 \Big( \big(\sin \theta  \sigma +\frac{2 e^{2 B } B_{,\theta} }{r}\big)du-d\theta \Big)^2\\
& -r^2 \sin^2\theta \Big(\big(\rho+\frac{2 e^{2 B } \csc ^2\theta  B_{,\phi} }{r}\big)du -d\phi \Big)^2,\\
V=&-\frac{1}{3} e^{2 B } \Lambda  r^3+2 \cos \theta  \sigma (u) r^2 -2M\\
&+e^{2 B } \left(4 B_{,\phi} ^2 \csc ^2\theta +2
   B_{,\phi \phi}  \csc ^2\theta +4 B_{,\theta} ^2+2 \cot \theta  B_{,\theta}
   +2 B_{,\theta \theta} +1\right) r.
\end{aligned}
\end{equation}
We call this family of solutions {the Bondi-Sachs rockets}. Physically,
$\sigma(u)$ results in the acceleration of the systems. Moreover, $\sigma(u)$, $\rho(u)$ result in the variation of the shear.

\subsection{Case (i): B=B$(u)$}

In this case (\ref{lastequ0}) implies that $M=M(u)$. Also we can set $B=0$ by the coordinate transformations
(\ref{coortr}). Using coordinate transformation $\phi\rightarrow\phi-\int \rho $, we can reduce
$\rho(u)=0$. So the metric becomes
\begin{equation}
\begin{aligned}
ds^2=&\Big(1-\frac{\Lambda}{3}r^2-\frac{2M(u)}{r}
+2\cos \theta \sigma r -r^2 \sigma^2 \sin^2 \theta\Big)du^2+2du dr \\
&+2r^2\sin \theta \sigma du d\theta -r^2\Big(d\theta^2+\sin^2 \theta d \phi ^2\Big). \label{Solution1}
\end{aligned}
\end{equation}
Moreover, (\ref{lastequ}) gives $\eta=\eta(u,\theta)$ and
\begin{eqnarray}
- M_{,u}-3 \cos \theta  M \sigma=\frac{\eta ^2}{2}.\label{Solution1-m0}
\end{eqnarray}
This is the standard Kinnersley rocket with $\Lambda$, which was also obtained in terms of the pure radiation
Robinson-Trautman spacetimes \cite{Podolsky08}.

\subsection{Case (ii): B=B$(u, \theta, \phi)$}

The complete classification of the metrics is not known yet. If $B=B(u, \theta)$, we can use the coordinate
transformation $\phi\rightarrow\phi-\int \rho $ to reduce $\rho(u)=0$. Moreover, (\ref{lastequ0}) implies
$M=M(u, \theta)$. So the metric (\ref{BS-roc}) reduces to the axi-symmetric Bondi-Sachs rockets
\begin{eqnarray}
\begin{aligned}
ds^2=&e^{2B}\frac{V}{r} du^2 +2e^{2B}dudr -r^2 \Big( \big(\sin \theta  \sigma +\frac{2 e^{2 B } B_{,\theta} }{r}\big)du-d\theta \Big)^2
-r^2 \sin^2\theta d\phi ^2,\\
V=&-\frac{1}{3} e^{2 B } \Lambda  r^3+2 \cos \theta  \sigma (u) r^2
+e^{2 B } \left(4 B_{,\theta} ^2+2 \cot \theta  B_{,\theta}+2 B_{,\theta \theta} +1\right) r-2 M \label{axi}
\end{aligned}
\end{eqnarray}
where $B(u,\theta)$ satisfies
\begin{eqnarray}
\begin{aligned}
&e^{4 B}
   \Big(16 \cot \theta  B_{,\theta} ^3+4 B_{,\theta \theta}^2-4 B_{,\theta} ^2
   \left(\cot ^2\theta -4 B_{,\theta \theta}\right)-B_{,\theta \theta}\cot^2 \theta\\
   &+2 \cot \theta  B_{,\theta \theta \theta}+B_{,\theta}  \big(\cot \theta  \left(\csc ^2\theta
   +16 B_{,\theta \theta}+2\right)+8 B_{,\theta \theta \theta}\big)+B_{,\theta \theta \theta \theta}\Big)\\
&+4 M B_{,u}-2 M_{,u}-6 \cos \theta  M \sigma-3\sin \theta M_{,\theta} \sigma =\eta ^2(u, \theta). \label{B}
\end{aligned}
\end{eqnarray}
As examples for demonstration, we provide two exact solutions. For the first solution, we take
\[
B=\frac{\ln \sin\theta}{2}, \quad M=\tau(u)\csc^2 \theta, \quad \eta ^2 =-2 \tau'(u)\csc^2 \theta
\]
for certain function $\tau'(u)\leq 0$. The corresponding metric is
\begin{equation}
\begin{aligned}
ds^2=&-\Big((\frac{\Lambda}{3}+\sigma^2) r^2 \sin ^2\theta +\frac{2\tau(u) }{r\sin\theta}\Big)du^2 +2 \sin \theta dudr\\
&+2 r \Big(\cos \theta +r \sigma (u) \sin \theta \Big) dud\theta
-r^2\Big(d\theta^2+\sin^2\theta d\phi^2 \Big). \label{zz}
\end{aligned}
\end{equation}
For the second solution, we take
\begin{eqnarray*}
\begin{aligned}
B=&\frac{\ln(C+\cos\theta)}{2},\quad M=\frac{\tau(u)}{(C+\cos\theta)^2},\\
\eta^2 =&-\frac{2\tau'(u)}{(C+\cos\theta)^2}-\frac{6\sigma(u)\tau(u)(C\cos\theta+\cos2\theta)}{(C+\cos\theta)^3}
\end{aligned}
\end{eqnarray*}
for constant $C>1$. The corresponding metric is
\begin{eqnarray}\label{zzz}
\begin{aligned}
ds^2=&g_{00}du^2+2 \big(C+\cos \theta \big)dudr+2 r \sin \theta  \big(r \sigma-1 \big)dud\theta\\
&-r^2\Big(d\theta^2 +\sin^2\theta d\phi^2 \Big), \\
g_{00}=&-(C+\cos \theta ) \Big(C (\frac{\Lambda}{3}r^2 -1)+\cos \theta (\frac{\Lambda}{3} r^2 -2 r \sigma+1)\Big)\\
&-\frac{2 \tau (u)}{r (C+\cos \theta )}-\sin ^2\theta  (r \sigma -1)^2.
\end{aligned}
\end{eqnarray}
Note that (\ref{zz}) is singular at $\theta =0, \pi$ but (\ref{zzz}) is a regular solution.

The Kinnersley-type rockets (type II) in \cite{Bonnor96} are
 \begin{eqnarray}
\begin{aligned}
ds^2=&\Big(K-2Hr-\frac{2M(u)}{r}-b^2 f r^2\Big)du^2 +2du dr +2 b r^2\sin \theta du d\theta \\
&-r^2\Big(\big(1+\sin^2\theta h\big)^{-1} d\theta^2+\big(1+\sin^2\theta h\big) \sin^2 \theta d \phi ^2\Big) \label{K2}
\end{aligned}
 \end{eqnarray}
where $h(u,\theta)>-1$, $f(u,\theta)$, $b(u)$, $K(u, \theta)$ and $b(u, \theta)$ are functions determined by $h$ \cite{Bonnor96}.
The Bondi C-metric in \cite{Bonnor90} is
 \begin{eqnarray}
\begin{aligned}
ds^2=&\Big(\frac{V}{r} -r^2 \sin^2\theta  \sigma ^2 L  \Big)du^2 +2dudr -2 r^2\sin \theta \sigma du d\theta \\
&-r^2\Big(L^{-1} d\theta^2+L \sin^2 \theta d \phi ^2\Big) \label{B-C} \\
V=&- r^2 \sigma \big(2 \cos\theta +6 M \sigma \cos^2 \theta \big)+r \big(1+6M \sigma \cos\theta \big)-2M
\end{aligned}
 \end{eqnarray}
where $\sigma$, $M$ are constant, and $L=1-2M \sigma \cot^2\theta \cos\theta$. The Bondi C-metric is of algebraic type D \cite{SKMHH03}.
In particular, if we require $L=1$, the Bondi C-metric will reduce to the Kinnersley rocket with zero mass ($M$=0) or to the Schwarzschild
metric ($\sigma$=0).

We shall see whether (\ref{BS-roc}) has the same algebraic type with (\ref{K2}) or (\ref{B-C}).
As (\ref{K2}) and (\ref{B-C}) represent spacetime metrics with vanishing cosmological constant, we set $\Lambda =0$
in (\ref{BS-roc}), in this case we can can take $B=0$, $\rho=0$. So (\ref{BS-roc}) becomes
\begin{eqnarray*}
\begin{aligned}
ds^2=&\Big(1-\frac{2M(u)}{r}
+2\cos \theta \sigma r -r^2 \sigma^2 \sin^2 \theta\Big)du^2+2du dr \\
&+2r^2\sin \theta \sigma du d\theta -r^2\Big(d\theta^2+\sin^2 \theta d \phi ^2\Big) \label{Solution1}
\end{aligned}
\end{eqnarray*}
which is the Kinnersley rockets (type D). 

\section{Energy Balance}

In this section we shall study the gravitational radiation of the photon rocket solution (\ref{BS-roc}).
The emitted pure radiation energy is defined as (e.g.\cite{vonderGonna98})
\begin{eqnarray*}
E(u) =\frac{1}{8\pi}\lim _{r \rightarrow \infty}\int _{S^2 _r} T_{00},
\end{eqnarray*}
and the Bondi mass is defined as \cite{Bondi, Sachs}
\begin{eqnarray*}
m(u)=\frac{1}{4\pi}\int _{S^2} M.\label{Bondi-em}
\end{eqnarray*}

In the case of zero cosmological constant $\Lambda=0$, the system will lose mass and radiate energy in general.
But it does not occur if the news functions $c$, $d$ vanish and, in this case, $m'(u)=0$ \cite{Bondi, Sachs}.
However, the situation becomes sophisticated when the cosmological constant $\Lambda \neq 0$.
(We plan to study why $E(u)$ and $m(u)$ could represent, conceptually, the emitted pure radiation energy
and the rest mass in a forthcoming paper.) If $c=d=B=0$, the vacuum Einstein field equations imply
\[
M_{,u}=\frac{\Lambda}{2}\Big(\cot \theta N+N_{,\theta}+\csc \theta P_{,\phi} \Big)-\Big(\sin \theta M_{,\theta}+3\cos \theta M \Big)\sigma - M_{,\phi}\rho.
\]
As $X$, $Y$ may be nonzero, the above equality shows $m'(u)$ may not be zero. The acceleration of coordinates and
the cosmological constant will result in varying of the Bondi mass.

For the case (i), i.e., photon rocket solution (\ref{Solution1}),
\[
E(u)=\frac{1}{4}\int_0^{\pi} \eta^2 (u, \theta) \sin\theta d\theta=-m ' (u).
\]
Therefore $m'(u) +E(u)=0$ and the emission of pure radiation energy balances the loss of the Bondi mass.
It means that all loss of mass of system is due to the photon radiation.
No gravitational radiation occurs in this case, even with $\Lambda \neq 0$. (This fact was also observed for $\Lambda = 0$
in \cite{vonderGonna98}.)

For the axi-symmetric Bondi-Sachs rocket (\ref{axi}) in case (ii), we have
\begin{eqnarray*}
\begin{aligned}
m'(u)+E(u)=&\int_0^\pi MB_{,u}\sin\theta d\theta \\
&+\Big\{\frac{e^{4 B}\csc\theta }{4}\Big[\sin\theta\Big(B_{,\theta \theta}\sin\theta+2B_{,\theta}^2 \sin\theta-B_{,\theta} \cos\theta\Big)
\Big]_{,\theta}\Big\} _0^\pi.  \label{balance}
\end{aligned}
\end{eqnarray*}
If the right side of above identity vanishes, the pure radiation energy and the loss of the Bondi mass will balance.
This is indeed the case for the metric (\ref{zzz}) where $B$ satisfies $B_{,u}=0$ and $B_{,\theta \theta}\sin\theta+2B_{,\theta}^2 \sin\theta-B_{,\theta} \cos\theta=0$. However, for most $B$, there is no such balance.

\bigskip

\noindent{\bf Acknowledgement}: The authors would like to thank the referees for many valuable
suggestions and comments. Partial financial support from
National Science Foundation of China (grants 10725105, 10731080, 11021091),
and Chinese Academy of Sciences is gratefully acknowledged.

\appendix
\section{main equations in Bondi-Sachs spacetime}

In this appendix, the addition of a subscript to a function denotes its differentiation with respect to the
corresponding coordinate ($u=0$, $r=1$, $\theta=2$, $\phi =3$).

\[
\begin{aligned}
\Eq (1) = &-\beta _{1}+\frac{1}{2} r \Big(\gamma_{1} ^2 \cosh ^2 2 \delta+\delta _{1}^2\Big),\\
\Eq (2) = &-\Big(r^4e^{-2\beta}(e^{2\gamma}U_1\cosh2\delta+W_1\sinh2\delta)\Big)_1\\
&+2 e^{2 \gamma } \csc  \theta  \Big(2 \delta _{3} \gamma_{1} (2\sinh ^2 2 \delta +1) -2 \gamma _{3} \delta _{1}
+(2 \gamma _{3} \gamma _{1}+\gamma _{13})\cosh  2 \delta  \sinh  2 \delta\\
&-\delta _{13}\Big) r^2+2 \Big(-(2 \cot  \theta  \gamma _{1}-2 \gamma _{2} \gamma_{1}+\gamma _{12}) \cosh ^2 2 \delta
-4 \delta_{2} \gamma _{1} \cosh  2 \delta \sinh  2 \delta\\
&-\frac{2 \beta _{2}}{r}+2\delta _{2} \delta _{1}+\beta _{12}\Big) r^2,\\
\Eq (3) = &-\Big(r^4e^{-2\beta}(e^{-2\gamma}W_1\cosh2\delta+U_1\sinh2\delta)\Big)_1\\
&+2 \csc  \theta  \Big((2 \gamma _{3} \gamma _{1}+\gamma_{13}) \cosh ^2 2\delta +4 \delta _{3} \gamma _{1} \cosh 2 \delta \sinh  2 \delta -\frac{2 \beta _{3}}{r}\\
&+2 \delta _{3} \delta _{1}+\beta_{13}\Big) r^2+2 e^{-2 \gamma } \Big(-2\delta_{2} \gamma _{1} (2 \sinh ^2 2 \delta +1) -2 \cot  \theta  \delta _{1}+2 \gamma _{2} \delta _{1}\\
&-(2 \cot  \theta  \gamma_{1}-2 \gamma _{2} \gamma _{1}+\gamma _{12})\cosh  2 \delta \sinh 2 \delta
-\delta _{12}\Big)r^2,\\
\end{aligned}
\]
\[
\begin{aligned}
\Eq (4) = &-V_1-\Lambda r^2 e^{2\beta}-\frac{1}{4} e^{-2 \beta } \Big(2 U_{1}W_{1} \sinh  2 \delta +(e^{2 \gamma } U_{1}^2+e^{-2 \gamma } W_{1}^2)\cosh 2\delta\Big) r^4\\
&+\frac{1}{2} \csc  \theta  \Big(4 W_{3}+r W_{13}\Big) r-e^{2 \beta +2 \gamma } \csc ^2 \theta  \Big((2\beta _{3} \delta _{3}+4 \gamma _{3} \delta _{3}+\delta_{33})\sinh 2\delta\\
&+(\beta _{3}^2+2 \gamma _{3} \beta _{3}+2 \gamma _{3}^2+2 \delta_{3}^2+\beta _{33}+\gamma _{33})\cosh  2 \delta \Big)
+2 e^{2 \beta } \csc  \theta  \Big((\beta_{3} \beta _{2}\\
&+2 \delta _{3} \delta _{2}+\beta_{23})\sinh 2 \delta +(\cot  \theta  \delta _{3}+\beta_{2} \delta _{3}-\gamma _{2} \delta _{3}+\beta_{3}\delta _{2}+\gamma _{3} \delta _{2}\\
&+\delta_{23})\cosh 2\delta\Big)-e^{2 \beta -2 \gamma } \Big((\beta
   _{2}^2+\cot  \theta  \beta _{2}-2 \gamma _{2} \beta _{2}+2 \gamma _{2}^2+2 \delta _{2}^2
-3 \cot  \theta  \gamma _{2}\\
&+\beta_{22}-\gamma _{22}-1)\cosh  2 \delta+(3 \cot  \theta  \delta_{2}+2 \beta _{2} \delta _{2}-4 \gamma _{2} \delta_{2}+\delta _{22})\sinh 2 \delta\Big)\\
&+\frac{1}{2}\Big(4\cot \theta  U+4 U_{2}+r \cot  \theta  U_{1}+r U_{12}\Big)r,\\
\Eq (5) = &-\Big((r\gamma)_{01}\cosh2\delta+2r(\gamma_0\delta_1+\delta_0\gamma_1)\sinh2\delta\Big)
+\frac{1}{8} e^{-2 \beta } \Big(e^{2 \gamma } U_{1}^2\\
&-e^{-2 \gamma}W_{1}^2\Big) r^3 +\csc  \theta  \sinh  2 \delta \Big(-2 W \delta _{3} \gamma _{1}+W_{3} \delta _{1}-2 W \gamma_{3} \delta _{1}\Big) r\\
&-\sinh  2 \delta \Big(2 U \delta_{2} \gamma _{1}-\cot  \theta  U \delta _{1}+U_{2} \delta _{1}
+2 U \gamma _{2} \delta_{1}\Big) r\\
&+\frac{1}{4} e^{2 \gamma } \csc  \theta  \Big(4 U_{3} \delta _{1}\cosh  2 \delta +(\frac{2 U_{3}}{r}+U_{13})\sinh  2 \delta \Big) r \\
&+\frac{1}{4} \cosh  2 \delta  \csc  \theta \Big(\frac{2 W_{3}-4 W \gamma _{3}}{r}
-2 \gamma _{1}W_{3}-2\gamma _{3} W_{1}+W_{13}-4 W \gamma _{13}\Big) r\\
&-\frac{1}{4} e^{-2 \gamma } \Big(\frac{2(W_{2}-\cot\theta  W)}{r}\sinh  2 \delta+4 \delta _{1} (W_{2}-\cot\theta  W)\cosh  2 \delta\\
&+(W_{12}-\cot  \theta  W_{1})\sinh  2 \delta\Big) r-\frac{1}{4} \cosh  2 \delta  \Big(-\frac{2 \cot  \theta
U}{r}+\frac{4 \gamma _{2} U}{r}\\
&+2 \cot  \theta  \gamma _{1} U+4\gamma _{12} U+\frac{2 U_{2}}{r}
-\cot  \theta  U_{1}+2 \gamma _{2}U_{1}+2 U_{2} \gamma _{1}+U_{12}\Big) r\\
&+2 V \gamma _{1} \delta _{1}\sinh 2\delta+\frac{1}{2}\Big(\frac{V \gamma _{1}}{r}+V_{1} \gamma _{1}+V \gamma _{11}\Big)\cosh 2 \delta\\
&-\frac{e^{2 \beta +2 \gamma } \csc ^2 \theta
   \left(\beta _{3}^2+\beta _{33}\right)}{2 r}+\frac{e^{2 \beta } \csc  \theta  \left(\delta
   _{3} \beta _{2}-\beta _{3} \delta _{2}\right)}{r}\\
&+\frac{e^{2 \beta -2 \gamma } \left(\beta _{2}^2-\cot  \theta  \beta
   _{2}+\beta _{22}\right)}{2 r},\\
\Eq (6) = &-\Big((r\delta)_{01}-2r\gamma_0\gamma_1\sinh2\delta\cosh2\delta\Big)\\
&+\frac{1}{8} e^{-2 \beta } \Big(2 U_{1} W_{1}\cosh 2 \delta+ (e^{2 \gamma } U_{1}^2+e^{-2 \gamma } W_{1}^2 )\sinh2 \delta \Big) r^3\\
&-\frac{1}{4} e^{2 \gamma } \csc  \theta  \Big(4 U_{3} \gamma _{1}
\cosh ^2 2 \delta +\frac{2 U_{3}}{r}+U_{13}\Big) r-\frac{1}{2} \csc  \theta\Big(\frac{2 W \delta _{3}}{r}\\
&+W_{1} \delta _{3}+2 (W_{3} \gamma _{1}-2 W \gamma _{3} \gamma _{1})\cosh 2\delta  \sinh  2 \delta+W_{3} \delta _{1}
+2W \delta _{13}\Big) r \\
&-\frac{1}{4} e^{-2 \gamma } \Big(-4 (W_{2}-\cot  \theta W) \gamma _{1} \cosh ^2 2 \delta +\frac{2 (W_{2}-\cot\theta  W )}{r}\\
&-\cot  \theta  W_{1}+W_{12}\Big)r -\frac{1}{2}
   \Big(\frac{2 U \delta _{2}}{r}+U_{1} \delta _{2}+\cot  \theta  U \delta _{1}+U_{2} \delta
   _{1}+2 U \delta _{12}\\
\end{aligned}
\]
\[
\begin{aligned}
& -2 (-\cot  \theta  U \gamma _{1}+U_{2} \gamma _{1}+2 U \gamma _{2} \gamma
   _{1})\cosh 2\delta  \sinh  2 \delta\Big) r \\
&+\frac{1}{2} \Big(-2 V \gamma _{1}^2 \cosh  2 \delta
   \sinh  2 \delta+\frac{V \delta _{1}}{r}+V_{1} \delta _{1}+V \delta _{11}\Big)\\
&-\frac{e^{2 \beta
   +2 \gamma } \csc ^2 \theta  (\beta _{3}^2+\beta_{33})}{2 r}\sinh  2 \delta
-\frac{e^{2 \beta -2 \gamma } (\beta _{2}^2-\cot  \theta  \beta _{2}+\beta _{22})}{2r}\sinh  2 \delta\\
&-\frac{e^{2 \beta } \csc  \theta  (\cot
 \theta  \beta _{3}-\beta _{2} \beta _{3}-\gamma _{2} \beta
   _{3}+\gamma _{3} \beta _{2}-\beta _{23})}{r}\cosh  2 \delta .
\end{aligned}
\]

\end{document}